\documentstyle[sprocl,epsf]{article}

\begin{document}

\title{A NEW PARAMETRIZATION OF POLARIZED PARTON DISTRIBUTION FUNCTIONS}

\author{T. Morii}

\address{Faculty of Human Development, Kobe University, Kobe 657, Japan,
\\E-mail: morii@kobe-u.ac.jp} 

\maketitle\abstracts{Requiring the positivity and counting rule, 
we propose a new parametrization of the polarized parton distribution 
which reproduce the data well.  
We found that the NLO analysis is absolutely necessary to extract the 
correct information on the polarized gluon distributions. 
The calculated results of $x\Delta d_v(x)$ and $\Delta d_v(x)/d_v(x)$ 
seem to be consistent with the SMC data at $Q^2=$10GeV$^2$.}

Recently there has been renewed interest on polarized
parton distributions in the nucleon, largely stimulated by the development 
of the next-to leading order(NLO) QCD calculations of polarized parton 
distribution functions\cite{Mer} and acquisition of high-precision 
polarized data on various targets\cite{Adam}.

In this work, requiring reasonable physical conditions, (1)the 
positivity condition and 
(2) the counting rule for the helicity dependent parton 
distribution functions, we propose 
a new parametrization\cite{Goto} of the polarized parton distribution function
which is more reliable and fits better to the data than other 
parametrizations proposed so far.

Based on the positivity condition and counting rule for helicity 
dependent distributions,
we propose the following 
functional form of the polarized parton distributions at an initial
scale $Q_0^2$,
\begin{equation}
\Delta f_i(x, Q^2_0) = A_i~x^{\alpha_i}~
(1 + \gamma_i~x^{\lambda_i})~f_i(x, Q^2_0)~,
\label{eqn:PPDF2}
\end{equation}
where $\Delta f_i(x, Q^2_0)$ and $f_i(x, Q^2_0)$ are polarized and 
unpolarized distributions with $i=u_v$, $d_v$,
$\bar q$ and $g$. $\alpha_i$, $\gamma_i$ and $\lambda_i$ are free 
parameters and $A_i$ is the normarization factor.  Further constraints
are obtained, 
\begin{equation}
A_i = \frac{1}{1+\gamma_i}~,~~~\alpha_i \geq 0~,
~~~\gamma_i \neq -1~,~~~\lambda_i > -\alpha_i~,
\label{eqn:PARS}
\end{equation}
from the condition, 
$f_i^+(x, Q^2_0)\gg f_i^-(x, Q^2_0)$
for $x\to 1$, derived from the counting rule assumption\cite{BBS95}
and a reasonable assumption that the first 
moment of each polarized
parton distributions should not become infinite at $Q_0^2$.
The remaining 3 parameters($\alpha_i$, $\gamma_i$ and 
$\lambda_i$) for each $i$ are determined from $\chi^2$--fitting 
to the data.  

In this analysis, we assume the SU(3)$_f$ symmetry and take 
$\Delta\bar u(x)=\Delta\bar d(x)=\Delta\bar s(x)$
at an initial $Q_0^2$, for simplicity.
Then the first moments of $\Delta u_v$ and
$\Delta d_v$, which are written as $\eta_{u_v}$ and $\eta_{d_v}$, 
respectively, can be described in terms of
weak hyperon decay constants $F$ and $D$ as follows,
$\eta_{u_v}-\eta_{d_v}=F + D$,
$\eta_{u_v}+\eta_{d_v}=3F - D$.
Using updated values of the $\beta$--decay data
of np, $\Lambda$p, $\Xi\Lambda$ and
$\Sigma$n, $F$ and $D$ are calculated as
$F = 0.463\pm0.008$,
$D = 0.804\pm0.008$
which lead to $\eta_{u_v}=0.926$ and $\eta_{d_v}=-0.341$.
Then, for $i=u_v$ and $d_v$, we can
reduce the number of free parameters from 3 to 2 by using 
those fixed first moments.
Thus, we are to have 10 parameters,
$\alpha_{u_v}, \gamma_{u_v},
\alpha_{d_v}, \gamma_{d_v},
\alpha_{\bar q}, \gamma_{\bar q}, \lambda_{\bar q},
\alpha_g, \gamma_g, \lambda_g$, which can be determined 
by the $\chi^2$--analysis to
the experimental data, $A_1(x, Q^2)$, of the pol--DIS on 
the proton, neutron and deuteron.

Using the GRV parametrization for unpolarized parton distribution 
functions at the LO and
NLO\cite{GRV95} and the $R_{1990}$ values\cite{Whitlow} for $R(x, Q^2)$, 
we have carried out the $\chi^2$--analysis to the data by minimizing
$\chi^2=\sum(A_1^{data}-A_1^{calc})^2/(\sigma_{A_1}^{data})^2$ with
the help of the CERN subroutine MINUIT, where $\sigma_{A_1}^{data}$
is the experimental error including both systematic and statistical errors.
The QCD parameter $\Lambda_{QCD}$ in the DGLAP equations
is taken as
$\Lambda_{LO}^{(3, 4, 5)} = 232, 200, 153 {\rm MeV}$ for the LO and 
$\Lambda_{NLO, \overline{MS}}^{(3, 4, 5)} = 248, 200, 131 {\rm MeV}$ 
for the NLO with $N_f=3$.
The results of $\chi^2$--analysis are substantially better for the 
NLO($\chi^2$/d.o.f=304.25/339) than for the LO($\chi^2$/d.o.f=368.84/339).
The obtained parameters reproduce well the experimental data and 
thus the model of our parametrization seems promising.
Since the difference between the NLO
($\overline{\rm MS}$--scheme) and LO is due to if
the polarized gluon contributes to the structure function $g_1$ via
Willson coefficient, the result that the fitting at the NLO is better 
than at the LO implies that the polarized gluon has a nonzero 
contribution to the nucleon 
spin, $i.e.$ $\Delta g\ne 0$ at $Q^2_0$.
In this analysis, we have also found that the fitting at the NLO
is more sensitive to the polarized gluon distribution than that at the LO.
Therefore we can say that the NLO analysis is absolutely necessary to extract
informations on the polarized gluon distribution. 
Details will be given in our forthcoming paper\cite{Goto}. 

\begin{figure}[t]
\begin{center}
\hspace{-0.5cm}
\epsfxsize=12cm
\epsfysize=6cm
\epsfbox{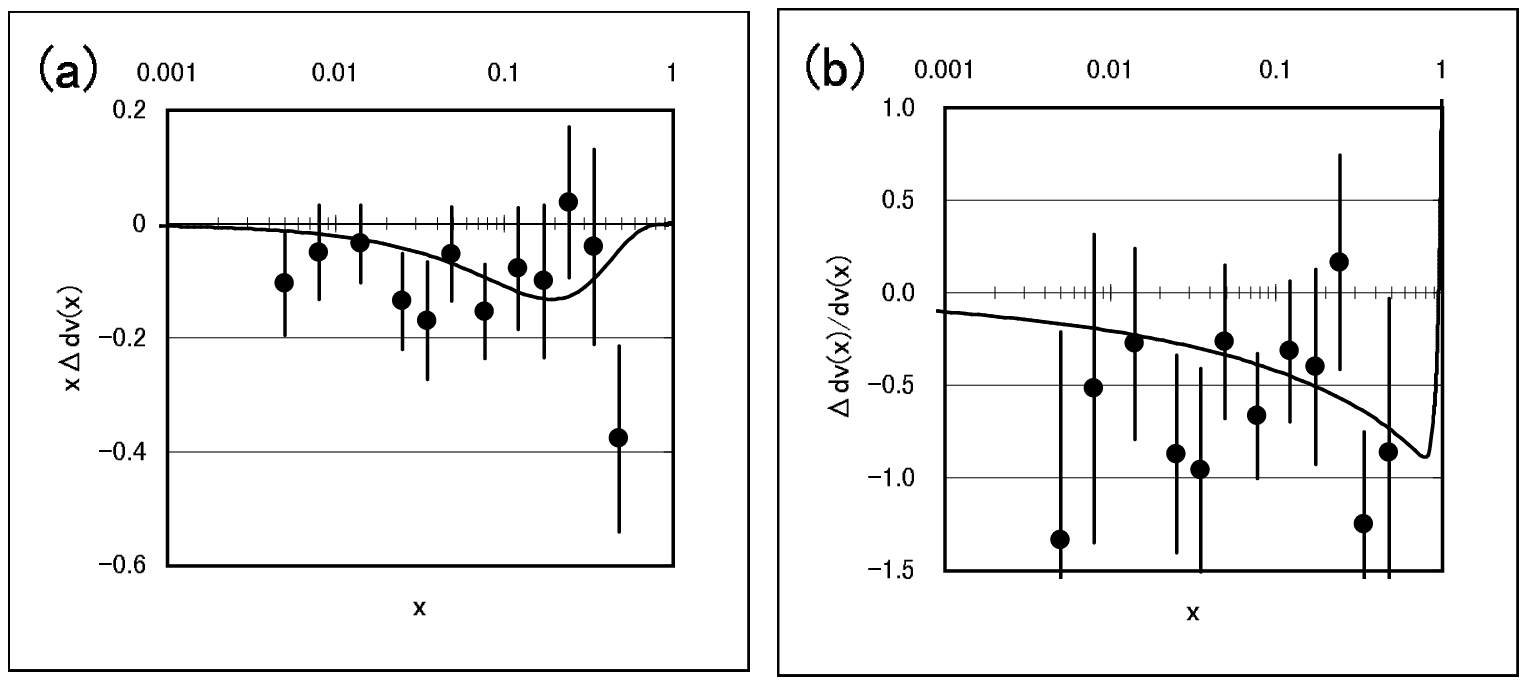}
\caption{Calculated distributions of $x\Delta d_v(x)$ and $\Delta 
d_v(x)/d_v(x)$ at $Q^2=$10GeV$^2$.  Data are from Ref.[7].
 \label{fig:xv}}
\end{center}
\end{figure}

As an additional test of the present parametrization, we have presented
our calculations of $x\Delta d_v(x)$ and $\Delta d_v(x)/d_v(x)$ 
and compared with the SMC data\cite{Adeva98} at $Q^2=$10GeV$^2$ in fig.1.  
It is interesting to 
see that the results seems to be consistent to the data, though the 
error bars of the data are large.

In this analysis, we have not taken into account the SU(3) flavor
symmetry breaking effect of the polarized parton distributions which
needs further investigation.

\section*{Acknowledgments}
This work was done in the collaboration with the RIHC-SPIN-J 
Working group members; Y. Goto(RIKRN), N. Hayashi(RIKEN), 
M. Hirai(Saga U.), H. Horikawa(Kobe U.), S. Kumano(Saga U.),
M. Miyama(Saga U.), N. Saito(RIKEN), T.-A. Shibata(TIT), 
E. Taniguchi(TIT) and T. Yamanishi(FUT).

\end{document}